\documentstyle[preprint,aps,epsfig]{revtex}

\tighten

\title{ Quantum analog of channeled electron trajectories in periodic 
       magnetic and electric fields }

\author{
     A. Manolescu$^{\mathrm{a}}$, S. D. M. Zwerschke$^{\mathrm{b}}$, M. Ni\c{t}\u{a}$^{\mathrm{a}}$, 
     U. J. Gossmann$^{\mathrm{b}}$, and R. R.\ Gerhardts$^{\mathrm{b}}$}
\address{
     $^{\mathrm{a}}$Institutul Na\c{t}ional de Fizica Materialelor, C.P. MG-7 
     Bucure\c{s}ti-M\u{a}gurele, Rom\^ania,\\
     $^{\mathrm{b}}$Max-Planck-Institut f\"ur Festk\"orperforschung, Heisenbergstra\ss e 1, 
     D-70569 Stuttgart, Federal Republic of Germany}

\begin{document}

\maketitle

\begin{abstract}
   We calculate the quantum states corresponding to the drifting and
   channeled classical orbits in a two-dimensional electron gas (2DEG)
   with strong magnetic and electric modulations along one spatial
   direction, $x$.  The channeled states carry high, concentrated currents
   along the $y$ axis, and are confined in an effective
   potential well. The quantum and the classical states are compared.\\

\end{abstract}
   
\vspace{2cm}

The classical equations of motion of an electron in a plane, in the
presence of a strong unidirectional magnetic or electric modulation,
plus an external perpendicular magnetic field, have two types of
solutions: drifting and channeled orbits.  
The former are slow, with self-intersecting cyclotron loops, while the
latter are fast, snaking around the lines of zero magnetic field in the
magnetic modulation and in the vicinity of the potential minima in the
electric one.  The channeled orbits are in the classical explanation responsible 
for the positive magneto resistance observed for low magnetic fields\cite{Be,MG}.
Recent experiments show a giant
effect in strong magnetic modulations, with estimated amplitude 0.6 T and
period 500 nm \cite{YWG2}.  Such conditions seem to require quantum mechanical
calculations.
Quantum states in a linearly varying magnetic field have been
calculated in \cite{Mu}.  

We consider the 2DEG, in the
$(x,y)$ plane, in a periodic magnetic field which has the form 
${\bf B}(x)=(B_0+B_1\cos\,Kx){\bf e}_z$.  The case with zero average
field has been recently discussed \cite{IP}.  Here we always keep
a finite $B_0$, denoting by $s=B_1/B_0$ the modulation strength.
The Hamiltonian is $H=({\bf p}+e{\bf A}(x))^2/2m$, where, in the
Landau gauge, ${\bf A}(x)=(0,B_0x+(s/K)\sin\,Kx)$.  
Classically, the possible trajectories are determined by two conserved
quantities, the energy $E$ and the canonical momentum in $y$-direction,
$p_y\equiv -X_0\omega_0m$.  Energy conservation, $E=mv_x^2/2+V(X_0;x)$,
can be used to calculate the velocity component $v_x(X_0,E;x)$
\cite{MG}.  Since $B_0\neq 0$ the electron is bound in $x$-direction, so
that the orbit has turning points $x_\pm$ given by the condition
$v_x(X_0,E;x_\pm)=0$.
Quantum mechanically, translational
invariance in $y$-direction implies wave functions of the form
$L_y^{-1/2}\exp(-iX_0y/l_0^2)\psi_{n,X_0}(x)$, where $L_y$ is a
normalization length, $n=0,1,2,...$, and $X_0$ is a
center coordinate.  We denote by $l_0=(\hbar/eB_0)^{1/2}$ the magnetic
length and by $\omega_0=eB_0/m$ the cyclotron frequency, corresponding
to the average field.  The electron effective mass is that of GaAs,
$m=0.067m_0$, and we assume spin degeneracy.

The reduced wave functions $\psi_{n,X_0}(x)$ are the solutions of a
one-dimensional Schr\"odinger equation with the effective potential 
$V(X_0;x)=\hbar\omega_0(x-X_0+{s\over K}\sin\,Kx)^2/2l_0^2$ \,.
For a fixed $X_0$ it has extrema at the positions where the magnetic 
field is zero, i.~e.\ at the roots of $1+s \cos\,Kx=0$, and additional minima 
at the 
points where $V(X_0;x)=0$.  We use the Landau wave functions (solutions 
for $s=0$) as the basis of our Hilbert space to obtain the eigenstates 
for $s\neq 0$ by numerical diagonalization.  The necessary basis size is here 
between 150-300 Landau levels.  
We can distinguish two situations:  $s<1$, when the magnetic field
$B(x)$ always points up, and $s>1$, when it has alternating 
sign.  For $s<1$ the energy spectra consist in more or less perturbed 
Landau levels reorganized in periodic Landau bands, $E_{n,X_0}$, with the
first Brillouin zone defined by $0\le X_0 < a\equiv 2\pi/K$.  The extension 
of the perturbed wave functions in the $x$ direction is about $2 R_{\mathrm{c}}$,
with $R_{\mathrm{c}}=l_0\sqrt{2n+1}$ the cyclotron radius in the
average field \cite{GMG}.

\begin{figure}
\epsfig{file=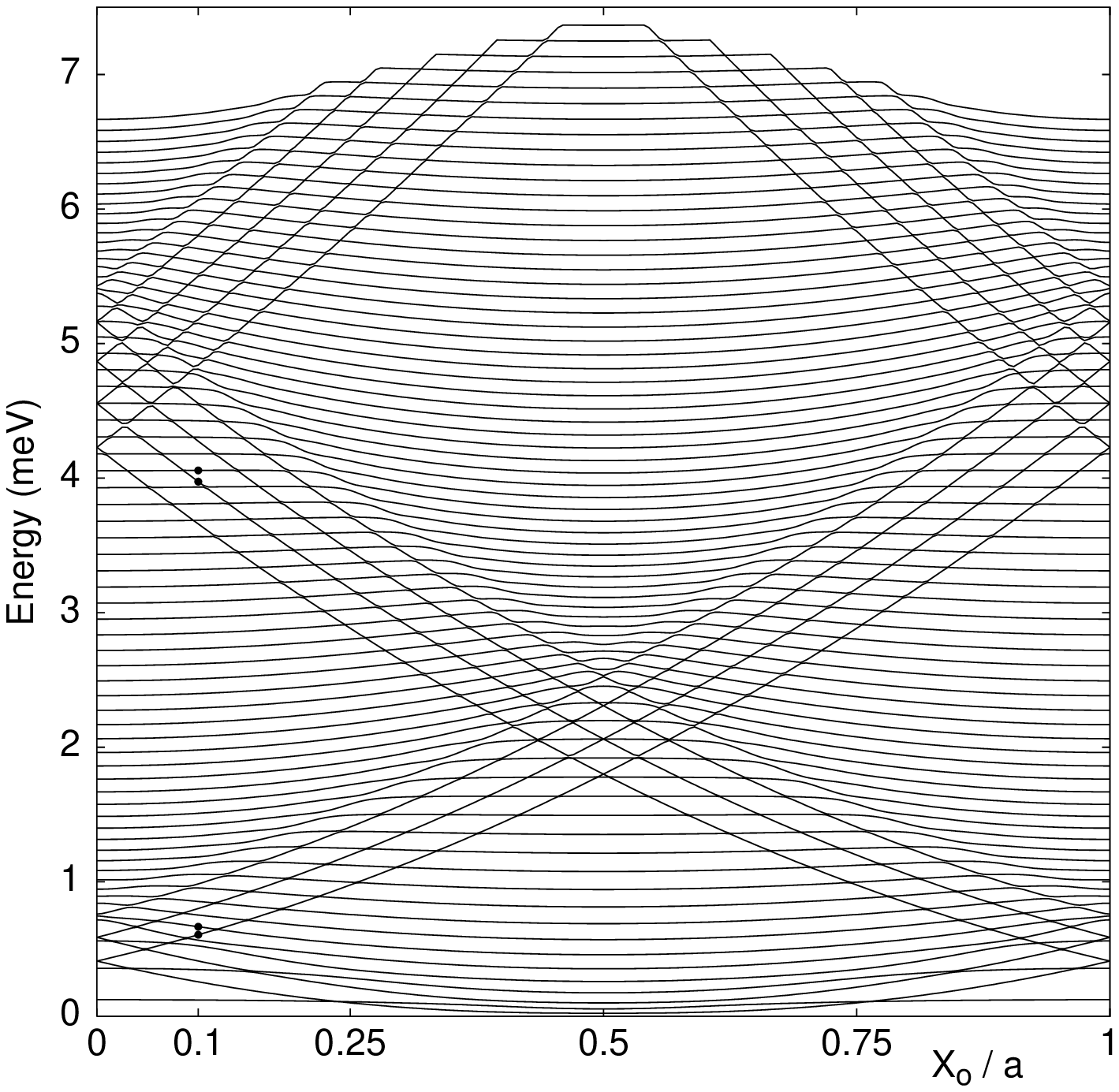}
\caption
{\label{pic1}Energy spectrum for $s=2$, $B_0=0.05$ T, $B_1=0.1$ T, 
and $a=800$ nm. The dots mark the states evaluated in Fig.~\ref{pic2}.}
\end{figure}

\begin{figure}
\epsfig{file=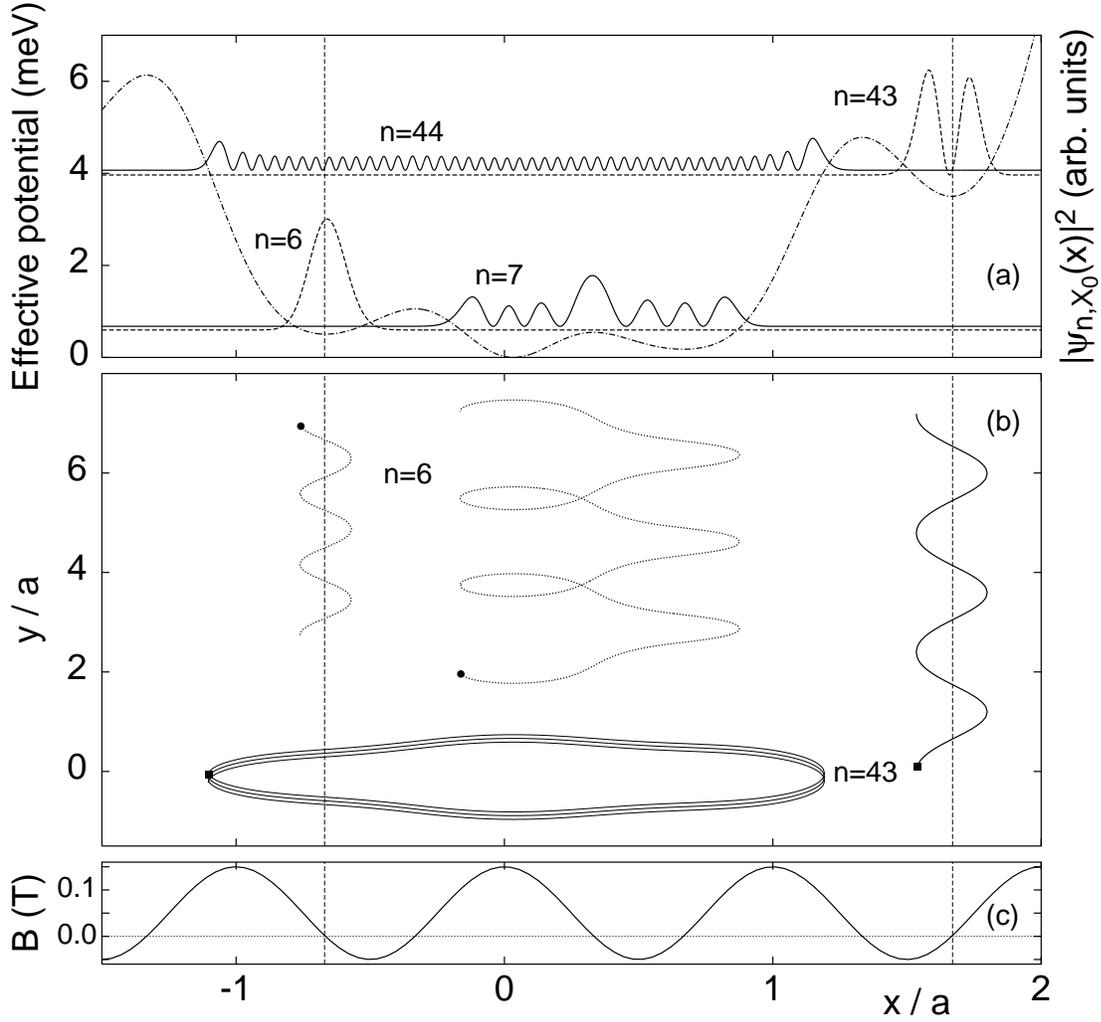}
\caption
{\label{pic2}Comparison of wave functions and classical trajectories for 
$X_0=0.1a$.
(a) The squared modulus of the wave functions and the corresponding effective 
potential (dash dotted line) for the states marked by the points in Fig.~\ref{pic1}.
The eigenvalues of the states are indicated by an energetical offset (left scale). 
(b) Trajectories for energies corresponding to $n=6$ and $n=43$,
beginning marked with dots.
(c) The periodic magnetic field.}
\end{figure}

For $s>1$, the energy spectra are more complicated, with strong overlap
of the bands, Fig.~\ref{pic1}.  To understand the nature of the states
we plot in Fig.~\ref{pic2}~(a) four characteristic states, indicated
with dots in Fig.~\ref{pic1}, having the quantum numbers $n=6, 7, 43,$
and $44$, 
and the same center coordinate
$X_0=0.1a$, so that all of them have the same effective potential.
Corresponding classical orbits are shown in Fig.~\ref{pic2}~(b).  There
are eight classical orbits, since for each of the four selected energies
there are {\em two} solutions.  We show only the four orbits with
energies $E_{6,X_0}$ and $E_{43,X_0}$, which are similar to those with
$E_{7,X_0}$ and $E_{44,X_0}$, respectively. Three cycles of the
classical orbits are plotted in each case.

\begin{figure}
\epsfig{file=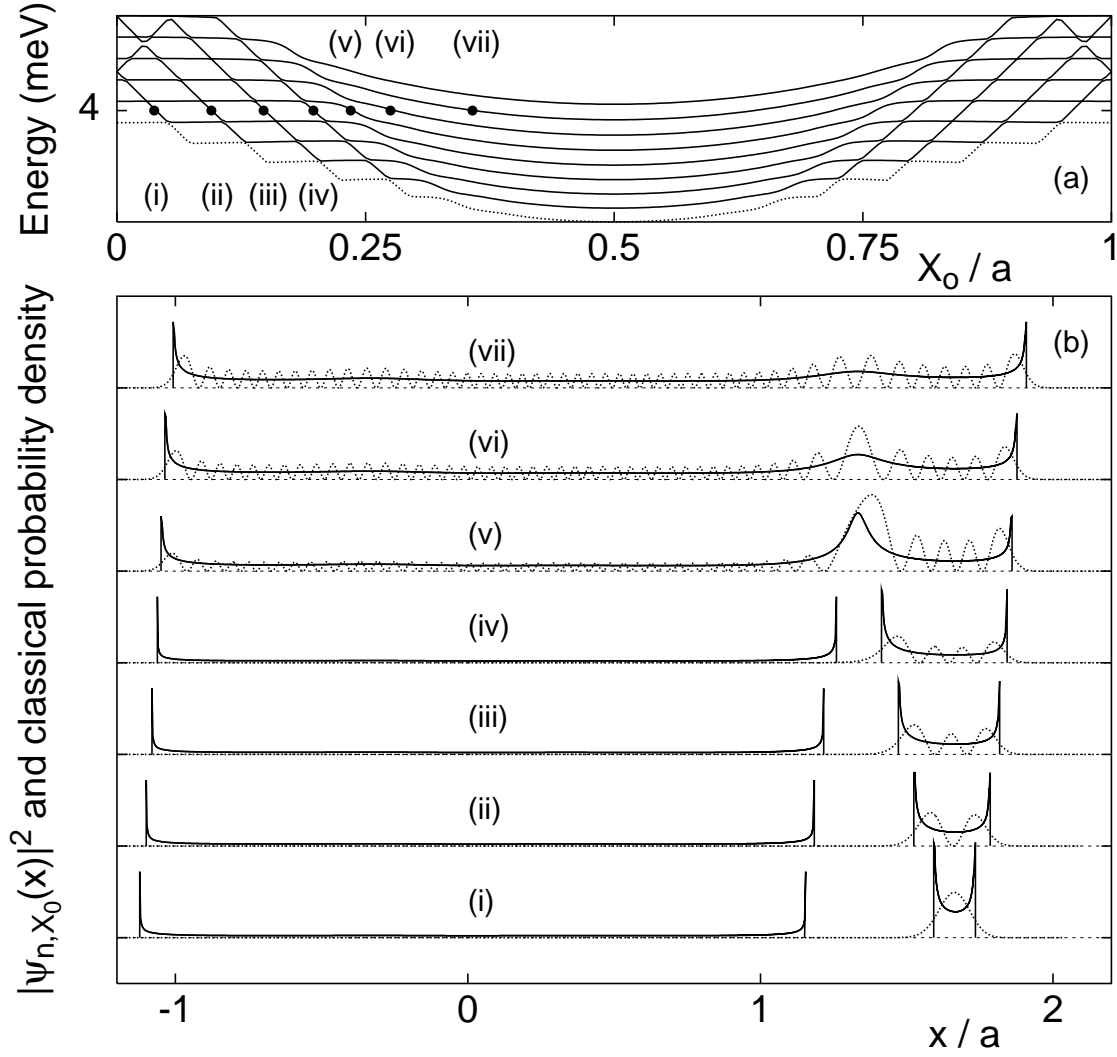}
\vspace{3cm}
\caption
{\label{pic3}
(a) An extract from energy spectrum, Fig.~\ref{pic1}. Dotted line:
band $n=41$.
(b) Quantum (dotted) and classical (solid lines) probability density at fixed 
energy, $E=4$ meV,
for the states marked in (a), in arbitrary units. 
Amplitudes of plots (v), (vi), and (vii) are amplified by a factor 5.}
\end{figure}

It is clearly seen, that the wide-spread quantum states, $n=7,\ 44$,
correspond to the drifting orbits of the classical picture. The states
$n=6,\ 43$, localized within shallow valleys of $V(X_0;x)$, correspond
to channeled orbits snaking along the lines with $B(x)=0$ \cite{Mu}.
The state with $n=6$ is the lowest (no
node) quantized state bound near $x/a=-2/3$, and that with $n=43$ the
second (one node) near $x/a=5/3$. 
The energy spacing between the quantum states depends on the width of
the potential well, hence it is larger for the `localized' channeled
states than for the `extended' drifting ones.  Also, hybridization
effects forbid degeneracy (for a fixed $X_0$) and hence the quantum
analogs of channeled and drifting orbits cannot have the same
energy, unlike the classical solutions.  
The apparent intersections of
the Landau bands are in fact anti-crossings, with extremely small gaps,
due to the exponentially small spatial overlap of nearly isoenergetic
states separated by large effective potential barriers. We indicated 
one individual band with the dotted line in
Fig.~\ref{pic3}~(a).  Now the interpretation of Fig.~\ref{pic1} becomes
clear: the drifting states give the branches with weak dispersion, and
the channeled states the branches with strong dispersion, i.e. with high
group velocity $\langle v_y\rangle=-(m\omega_0)^{-1}dE_{n,X_0}/dX_0$,
back-folded like free electron parabolas.

The motion in $x$-direction being periodic, one
can define a classical localization probability density, $W(x)$, by
comparing the time $dt$, which the electron spends in the interval $dx$
at $x$, with the period $T$: $W(x)dx=dt/T$. Due to the vanishing
velocity this function diverges at $x_\pm$.  In Fig.~\ref{pic3} we
compare the quantum mechanical probability density with the classical
one at the energy $E=4$ meV, for which we have 14 states in the Brillouin
zone. Due to the symmetry we display only the states with $X_0<a/2$,
indicated with dots in the relevant part of the spectrum,
Fig.~\ref{pic3}~(a).  The wave functions of the first four states (with
different $X_0$), belonging to energy-band segments with strong
dispersion, are now located in minima of different effective potentials,
but are channeled around the same line $B(x)=0$.  At the chosen energy
the classical trajectory is split into channeled and drifting orbits.
The classical solutions merge for the states (v)-(vii), which in the
quantum description belong to those regions of the spectrum where only
drifting states are possible, widely spread in the whole effective
potential.

For $s$ bigger than in Fig.~\ref{pic1}, when $V(X_0;x)$ may 
have more than one zero, the drifting states become more complicated.
This situation and other details will be discussed in a forthcoming paper.

\begin{figure} 
\epsfig{file=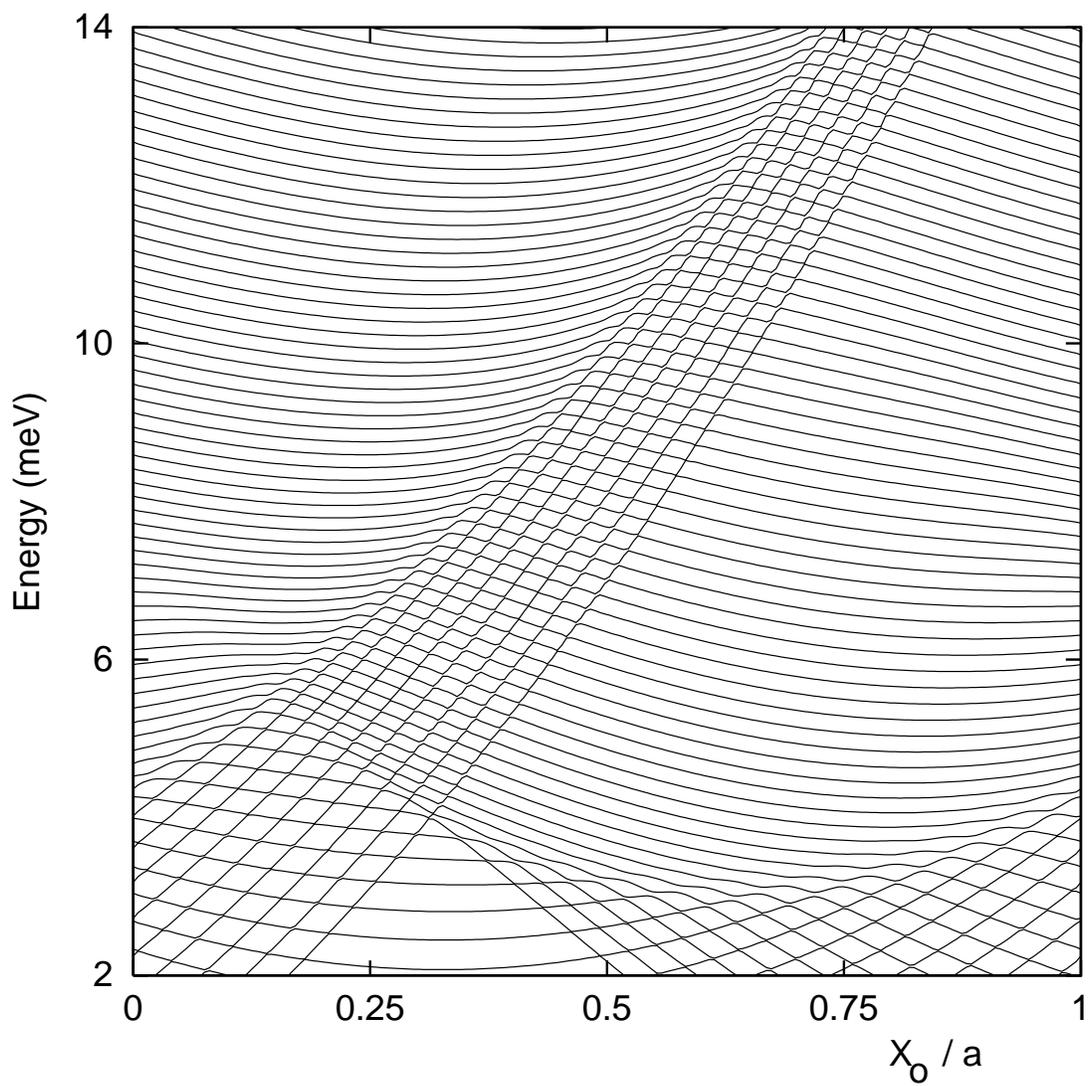}
\caption
{\label{pic4}Energy spectrum for combined magnetic and electric modulations, 
shifted with $\pi/2$.  $B_0=0.1$ T, $B_1=0.08$ T, and $U=3$ meV.} 
\end{figure}

In the modulated systems fabricated by the deposition of metallic
micro-strips of magnetic material on top of the semiconductor there is
always a stress-induced electric modulation which may be shifted in phase with
respect to the magnetic one \cite{YWG1}.  In a simple sinusoidal model
for the electric modulation the effective potential becomes
$V(X_0;x)+U\cos Kx$, where $U$ is the electrostatic potential amplitude.
For a pure electric modulation ($B_1=0$), when $U/\hbar\omega_0$ is
sufficiently large, the effective potential has local minima situated
close to the minima of the real potential where again channeled states
are captured. The physical mechanism is now a combined effect of
strong electric and weak Lorentz forces  
and the channeled orbits vanish with increasing energy \cite{Be}.
The channeled states carry positive and negative currents along the $y$
axis which in Fig.~\ref{pic1} cancel each other within each 
unit cell.  For an asymmetric 
situation, resulting by combining phase-shifted
magnetic and electric modulations, 
it is possible to obtain channeled currents running only one
way along the $y$ axis, and never in the opposite direction, in a wide
energy interval. 
Of course, in an equilibrium situation there can be no net current, and
the current carried by channeled states is compensated by the current
 carried by the drifting states.  In Fig.~\ref{pic4} we have chosen a
 magnetic modulation with $s<1$ and an electrostatic potential
$U\cos(Kx+\pi/2)$.  The channeled states appear thus due to the electric
modulation.

In conclusion we have discussed the quantum drifting and channeled
electronic states in strong modulations and compared them with classical electron 
trajectories. For understanding the recent
experiments \cite{YWG2} quantum transport calculations in this regime could 
be of great interest.

%
This work was supported by the BMBF grant No. 01 BM 622.


\end{document}